\documentclass[twocolumn,twoside,slac_two]{revtex4}
\usepackage{graphicx}
\usepackage{fancyhdr}
\def\be{\begin{equation}}
\def\ee{\end{equation}}
\def\bea{\begin{eqnarray}}
\def\eea{\end{eqnarray}}

\def\greaterthansquiggle{\raise.3ex\hbox{$>$\kern-.75em\lower1ex\hbox{$\sim$}}}
\def\lessthansquiggle{\raise.3ex\hbox{$<$\kern-.75em\lower1ex\hbox{$\sim$}}}
\def\gl{\raise.3ex\hbox{$<$\kern-.68em\lower1ex\hbox{$>$}}}
\newcommand{\gts}{\greaterthansquiggle}
\newcommand{\lts}{\lessthansquiggle}
\newcommand{\cp}{\mbox{$\not \hspace{-0.15cm} C\!\!P \hspace{0.1cm}$}}
\def \lsim{\mathrel{\vcenter
     {\hbox{$<$}\nointerlineskip\hbox{$\sim$}}}}
\def \gsim{\mathrel{\vcenter
     {\hbox{$>$}\nointerlineskip\hbox{$\sim$}}}}

\begin{document}

\title{Moderately light charged Higgs in \cp  MSSM and NMSSM}

%

\author{Rohini M. Godbole}
\affiliation{Centre for High Energy Physics, Indian Institute of Science, 
Bangalore, 560012, India.}

\begin{abstract}
In this talk I discuss some aspects of the phenomenology of a moderately
light charged Higgs ($H^\pm$) with a mass $\gts 130$ GeV, lighter 
than the top quark,  at the LHC.  A charged Higgs in this mass range
is still allowed in next-to-minimal Supersymmetric Standard Model (NMSSM)
at low $\tan \beta$ as well as in CP-violating (\cp) Minimal Supersymmetric
Standard Model (MSSM)
for a certain choice of \cp parameters, still respecting all the LEP-II 
bounds. In both the cases, the $H^\pm$ has a large branching ratio in the 
$W^\pm \phi$ channel, where $\phi$ denotes a generic Higgs which 
is dominantly pseudoscalar and hence may 
be substantially lighter than the LEP-II mass bound. This $\phi$  
decays dominantly into a $b \bar b$ pair. Thus production of $H^\pm$ in the top
decay gives a striking 
$t \bar t$ signal at the LHC, where one of the top quarks decays into 
the $bb \bar b W$ channel, via $t \to b H^\pm, H^\pm \to \phi$ and 
$\phi \to b \bar b$. The characteristic correlation between the 
$b \bar b$, $b \bar b W$ and $b b \bar b W$ invariant mass peaks helps 
reduce the Standard Model (SM) background very effectively. For these low values of $\tan \beta$ 
the $H^\pm \rightarrow \tau \nu_\tau  $ channel does not provide any reach 
for the $H^\pm$. Thus this is a signal for both a light charged $H^\pm$ and a
light $\phi$, which is mostly  pseudoscalar in nature  and decays dominantly 
into a
$b \bar b$ pair.
\end{abstract}

\maketitle

\thispagestyle{fancy}
\section{\label{intro}Introduction}
In this talk I discuss some aspects of the phenomenology of a moderately
light charged Higgs with  $130\ ~\lts ~\ M_{H^\pm}~ \ \lts~m_t$ GeV, at the LHC.  
A charged Higgs in this mass range is still allowed in the NMSSM at low 
$\tan \beta$~\cite{Drees:1998pw,Accomando:2006ga}
as well as in \cp\ MSSM in the CPX-scenario~\cite{Pilaftsis:1999qt,
Accomando:2006ga}, respecting all the LEP-II bounds~\cite{Carena:2000ks,
Abbiendi:2004ww,Schael:2006cr,Accomando:2006ga}. In both cases, there exists a 
neutral Higgs boson, $\phi$, which is predominantly  pseudoscalar  and hence 
can be much 
lighter than the LEP-II bound of $90$ GeV in the CP-conserving 
MSSM~\cite{Schael:2006cr}.  As a result, there exists a small
window in the $\tan\beta$--$M_{\phi}$ plane at low $\tan \beta\
(3\ \lts~\ \tan \beta~ \ \lts~5)$  and
$M_\phi < 50$ GeV ~\cite{Carena:2000ks,Abbiendi:2004ww,Schael:2006cr,Accomando:2006ga}, 
which is still allowed in the aforementioned CPX scenario, even after 
LEP-II constraints are taken into account.  In both the cases, the NMSSM 
as well as the \cp MSSM,  the $H^\pm$ has a large branching ratio in the 
$W^\pm \phi$ channel~\cite{medpcpnsh} and the light $\phi$ 
decays dominantly into a $b \bar b$ pair.  This thus gives a striking 
$t \bar t$ signal at the LHC, where one of the top quarks decays into 
the $bb \bar b W$ channel, via $t \to b H^\pm, H^\pm \to \phi$ and 
$\phi \to b \bar b$. The characteristic correlation between the 
$b \bar b$, $b \bar b W$ and $b b \bar b W$ invariant
mass peaks helps reduce the SM background very effectively.
Further, in the CPX-scenario  this $\phi$ has reduced couplings
to a $t \bar t$ pair as well. As a result, in the above mentioned window
in the parameter space, none of the usual search channels for the neutral
Higgs will have a reach either at the Tevatron or at the 
LHC~\cite{Carena:2002bb,schumacher}. The decay of a
light $H^\pm$, produced in the $t$ decay, provides an additional
channel to search for such a neutral scalar and help `fill' this 
hole which exists in this case.  For these low values of $\tan \beta$, the
$H^\pm$ can not be searched via $H^\pm \rightarrow \tau \nu_\tau $ and thus
there is no LHC reach for the $H^\pm$ as well. Thus the suggested topology can 
provide  a signal for the both the light charged $H^\pm$ and the 
light $\phi$ (which is mostly a pseudoscalar) in these regions of 
the parameter space, both for the NMSSM and  the ~~\cp MSSM.

\section{\label{nmssmsec} Light $H^\pm$ in NMSSM.}
In the MSSM $\mu$ is stabilized at the EW scale `naturally' if one introduces 
an additional chiral superfield~\cite{mybook}, thus solving the so 
called $\mu$ problem.  This class of Supersymmetric models, with additional 
singlet Higgs fields, is called the next-to-minimal supersymmetric standard 
model (NMSSM)~\cite{Nilles:1982dy,Nilles:1982mp,mybook}.
Further,  the `little hierarchy' problem caused by
the non observation of the light neutral Higgs at LEP-II, may also be eased
in the NMSSM~\cite{Dermisek:2005ar}. 
While adding a singlet complex scalar field does not affect the charged Higgs
boson pair $H^\pm$ of the MSSM directly, it has a strong indirect
effect on the phenomenology of the $H^\pm$ boson.  This effect comes from the
modification of the MSSM mass relations between the doublet scalars
$H^0_{1,2}$ and pseudoscalar $A_1^0$ and the resulting modification
of the $H^0_1$ mass bound.  Because of this modification it is
possible to satisfy the LEP-2 limit on the $H^0_1$ mass even with a
light $A_1^0$, which in turn implies a moderately light $H^\pm$
boson.  The relaxation of the $A_1^0$ and $H^\pm$ mass limits of the
MSSM is most pronounced in the moderate $\tan\beta$ region.
The superpotential of the NMSSM Higgs sector in terms of the singlet and 
doublet Higgs superfields, $\hat{S}$ and $\hat{H}_1, \hat{H}_2$ 
respectively, can be written as~\cite{Nilles:1982dy}:
\be
W = \lambda \hat{S} \hat{H}_1 \hat{H}_2 - {k \over 3} \hat{S}^3.
\label{one}
\ee
The resulting $F$ term of the Higgs potential is
\be
V_F = \lambda^2 x^2 (v^2_1 + v^2_2) + \lambda^2 v^2_1 v^2_2 + k^2x^4 -
2\lambda k x^2 v_1 v_2,
\label{two}
\ee
where $x = \langle S \rangle$, $v_{1,2} = \langle H^0_{1,2}\rangle$
and $\tan\beta = v_2/v_1$.  The $D$-term is the same as in MSSM, i.e.
\be
V_D = {1\over8} (g^2 + g^{\prime 2}) (v^2_1 - v^2_2)^2 + {1\over2} g^2
v^2_1 v^2_2.
\label{three}
\ee
comparing the $F$-term of the MSSM, $V_F = \mu^2 (v^2_1 + v^2_2)$,
with Eq. \ref{two} one gets
\be
\mu = \lambda x \equiv \mu_{eff},
\label{four}
\ee
which naturally explains why the supersymmetric $\mu$ parameter has a
relatively low value as required for EWSB.  In fact this solution to
the so called $\mu$-problem of the MSSM was the original motivation for
the NMSSM.  But the remaining terms of Eq. \ref{two} lead to
modifications of the neutral Higgs boson masses of the MSSM.  In
particular the resulting upper bound of the lightest Higgs scalar mass is 
~\cite{Drees:1988fc,King:1995vk,King:1995ys,Ellwanger:1991bq,Pandita:1993tg,Elliott:1993bs},
\be
M^2_{H_1} \leq M^2_Z \cos^2(2\beta) + {2\lambda^2 M^2_W \over g^2}
\sin^2(2\beta) + \epsilon,
\label{five}
\ee
where the first term coming from the $V_D$ (Eq. \ref{three}) and the
radiative correction $\epsilon$ are the same as in MSSM.  But the
additional contribution in the middle comes from the second term of
$V_F$ (Eq. \ref{two}).

Note that this additional contribution is most pronounced in the low
to moderate $\tan\beta$ region, where the MSSM mass bound coming from
the first term of Eq. \ref{five} is very small.  Therefore it
relaxes the MSSM bound on  $M_{H_1}$  and hence the resulting lower limit 
on $M_A$ most
significantly over this range of $\tan\beta$.  This in turn relaxes
the lower limit of the charged Higgs mass, which is related to the
doublet pseudoscalar mass via
\be
M^2_{H^+} = M^2_A + M^2_W \left(1 - {2\lambda^2 \over g^2}\right)
\label{six}
\ee
along with a small radiative correction.  This is helped further due
to the additional (negative) contribution in Eq. \ref{six}.  Note that the
additional contributions of Eqs. \ref{five} and \ref{six} depend
only on the $\hat{S} \hat{H}_1 \hat{H}_2$ coupling $\lambda$, represented
by the first term of the superpotential of Eq. \ref{one}.  Therefore
the Eqs. \ref{five} and \ref{six} hold also for the so called
minimal nonminimal supersymmetric standard model (MNSSM), which
assumes only the first term of the 
superpotential~\cite{Panagiotakopoulos:1999ah,Panagiotakopoulos:2001zy,Panagiotakopoulos:2000wp}.

\begin{figure}
 \includegraphics*[scale=0.7,angle=0]{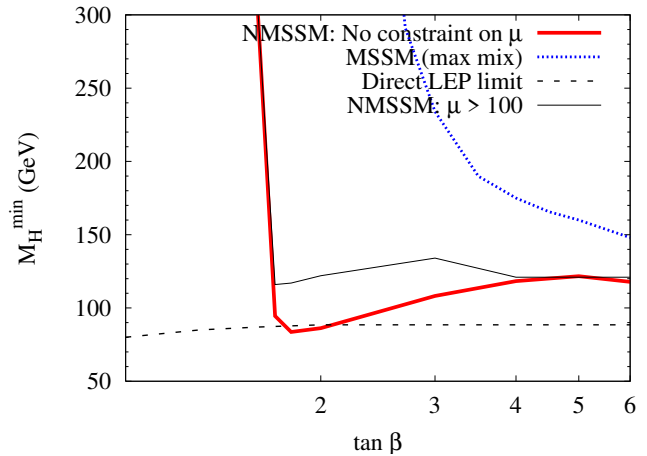}
 \caption{\label{fig:nmssmlim}The indirect lower bounds on the charged Higgs 
boson mass following from the LEP limits on the neutral Higgs bosons in the 
MSSM  with Maximal Stop Mixing (dotted curve), the NMSSM with no constraint
on $\mu$ (thick solid curve) and NMSSM with $\mu > 100$ GeV (thin solid curve). 
The direct LEP limit on the charged Higgs boson mass is also shown for 
comparison by the dashed curve.}
 \end{figure}

Finally the upper bound of Eq. \ref{five} will only be useful if one can 
find an upper limit on $\lambda$.  Such a limit can be 
derived~\cite{Drees:1988fc,King:1995vk,King:1995ys}  from the
requirement that all the couplings of the model remain perturbative
upto some high energy scale, usually taken to be the GUT scale.  Such
an upper limit on $\lambda$ has been estimated in~\cite{Drees:1998pw}
as a function of $\tan\beta$ using two-loop renormalization group equations.
For quantitative evaluation of the NMSSM Higgs spectrum  we consider the
complete Higgs potential as given in terms of these parameters in
~\cite{Bastero-Gil:2000bw}.  The lower limit of the $H^\pm$ mass has been
estimated as a
function of $\tan\beta$ in~\cite{Drees:1998pw} by varying all these five NMSSM
parameters over the allowed ranges, which include the constraints from
LEP-2.  The resulting $H^\pm$ mass limit is shown in Figure~\ref{fig:nmssmlim}
by the dark solid curve along with the most conservative MSSM limit
(shown by the dotted curve), corresponding to maximal stop
mixing, which gives the largest radiative correction $\epsilon$.  The
NMSSM limit has practically no sensitivity to stop
mixing. In addition the thin solid curve indicates the  limit in the NMSSM
where we require that the effective $\mu$ parameter given by Eq.~\ref{four}
be bigger than $100$ GeV as favoured by the chargino mass constraint from
LEP-II. The LEP-2 mass limit from direct search of 
$H^+ \rightarrow \tau^+ \nu$ events is also shown by the dashed curve for 
comparison~\cite{Yao:2006px}.
In fact it might be interesting to check whether the flavour
physics constraints allow this moderately light charged Higgs in this
region of the parameter space. There is no limit from Tevatron in the 
moderate $\tan\beta$ region shown in
Figure~\ref{fig:nmssmlim}.

One sees from Figure~\ref{fig:nmssmlim} that even the most conservative MSSM
limit implies $H^\pm$ mass $\geq 150$ GeV (175 GeV) for $\tan\beta \leq 6 \
(4)$.  In contrast in the NMSSM one can have a $H^\pm$ mass $\lsim
120$ GeV over this moderate $\tan\beta$ region, going down to the
direct LEP-2 limit of 86 GeV at $\tan\beta \simeq 2$.  Note however
that  requiring that  the effective $\mu$ parameter
$\mu_{eff} = \langle S \rangle \lambda$ be greater than
100 GeV, as favored by the LEP chargino search, increases this mass limit
to $\gsim 120$ GeV~\cite{Panagiotakopoulos:2001zy}.  The steep
vertical rise at left reflects the well-known fixed-point solution at
$\tan\beta = 1.55$, where the top Yukawa coupling blows up at the GUT
scale.  Thus allowing for possible intermediate scale physics one can
evade the steep NMSSM mass limit at low $\tan\beta$ ~\cite{Masip:1998jc}.
In contrast the MSSM limit holds independent of any intermediate scale physics
ansatz.

We have investigated the neutral scalar and pseudoscalar Higgs
spectrum of the NMSSM, when the $H^\pm$ lies near its lower mass limit
$(M_{H^+} \simeq 120 \ {\rm GeV})$.  The lightest scalar is dominantly
singlet $(M_{H_1} \simeq 100 \ {\rm GeV})$, while the doublet
scalars are relatively heavy $(M_{H_{2,3}} > 120 \ {\rm GeV})$.  On the
other hand there is often a light pseudoscalar $(M_{A^0_1} \simeq 50 \
{\rm GeV})$ with a very significant doublet component.  Consequently a
light charged Higgs boson of mass $\simeq 120$ GeV is expected to
decay dominantly via the standard $H^+ \rightarrow \tau^+ \nu$ mode.
Thus one can probe this mass range via the $t \rightarrow bH^+
\rightarrow b \tau^+ \nu$ channel at Tevatron and especially at
the LHC.
On the other hand a somewhat heavier charged Higgs boson $(M_{H^\pm} >
130 \ {\rm GeV})$ can dominantly decay via the $H^+ \rightarrow W^+ \
A_1^0$ channel~\cite{Drees:1999sb}.  In fact this seems to be a very
favorable channel to probe for not only $H^+$ but also a light
$A_1^0$ in the moderate $\tan\beta$ region, where the
$A_1^0$ is expected to decay mainly in to the $b\bar b$  mode. 
Table~\ref{tablenmssm} shows some illustrative samples of
NMSSM Higgs spectra where $H^+$ decays dominantly into the $W^+ A_1^0$ mode.
These results are obtained by scanning the NMSSM parameter space.  Note
that in each case the effective $\mu$ parameter
$\mu_{eff} = \lambda \langle S \rangle$ is greater than
100 GeV as favored by the LEP chargino limit.  The decay branching fractions
are shown along with the Higgs boson masses and the other model
parameters in Table~\ref{tablenmssm}. We see indeed that $B(H^\pm \rightarrow
A_1^0 W^\pm ) $ can be substantial.
\begin{table}
     \caption{\label{tablenmssm}
      Examples of dominant $H^\pm \rightarrow W A^0_1$
     decay in the NMSSM.  These decay branching fractions
    are shown along with the Higgs boson masses and the other model
     parameters.}
     \begin{tabular}{|c|c|c|c|c|c|}
     \hline
     $\tan\beta$ & $M_{H^+}$ & $M_{A^0_1}$ & $B_{A^0_1}$ &
     $\lambda,\kappa$ & $x=v_s/\sqrt{2},A_\lambda,A_\kappa$ \\
     & (GeV) & (GeV) & $(\%)$ & & (GeV) \\
     \hline
     &&&&& \\
     2 & 147 & 38 & 94 & .45,-.69 & 224,-8,2 \\
     &&&&& \\
     3 & 159 & 65 & 83 & .33,-.70 & 305,40,38 \\
     &&&&& \\
     4 & 145 & 48 & 89 & .28,-.70 & 563,170,85 \\
     &&&&& \\
     5 & 150 & 10 & 91 & .26,-.54 & 503,109,38 \\
     &&&&& \\
        \hline
    \end{tabular}
    \end{table}

\section{\label{cpvsec}Light $H^\pm$ in CP-violating MSSM}
Interestingly one can have a similar signal in the CP violating MSSM
due to large scalar-pseudoscalar mixing. In this case, the $h,H$ and $A$ of the MSSM are no longer mass eigenstates, even if we start with a tree level scalar
potential which is CP-conserving. Loop effects mix them and the mass eigenstates
$H_1, H_2$ and $H_3$ (ordered according to their masses) no longer have a 
definite CP\footnote{$H_1$ here is synonymous with $\phi$ of earlier discussion
in the Introduction.}.
The CP-violating MSSM allows existence of a light neutral Higgs boson ($M_{H_1}
\lsim 50$ GeV) in the CPX scenario in the low $\tan \beta (\lsim 5)$ region,
which could have escaped the LEP searches due to a strongly suppressed 
$H_1 Z Z$  coupling. The light charged $H^+$ decays dominantly into
the $W H_1$ channel giving rise to  a striking $t \bar t$ signal at the LHC,
where one of the top quarks decays into the $bb \bar b W$ channel, via
$t \to b H^\pm, H^\pm \to W H_1$ and $H_1 \to b \bar b$. The characteristic
correlation between the $b \bar b$, $b \bar b W$ and $b b \bar b W$ invariant
mass peaks helps reduce the SM background, drastically~\cite{Ghosh:2004cc}.
Note that this signal is identical to the NMSSM case discussed above.
 As already mentioned, a combined analysis of all the LEP results,
shows that a light neutral Higgs is still allowed in the CPX~\cite{Pilaftsis:1999qt} scenario in the CPV-MSSM.  The experiments provide
exclusion regions in the $M_{H_1}- \tan \beta $ plane for different values of
the CP-violating phase, with the various  parameters taking the following
values:
\bea
{\rm Arg}A_t = {\rm Arg}A_b = {\rm Arg}M_{\tilde g}=\Phi_{\rm CP},\\
M_{\rm SUSY}= 0.5~{\rm TeV}, M_{\tilde g} = 1~ {\rm TeV},\\
M_{\tilde B} = M_{\tilde W}= 0.2~ {\rm TeV},\\
\Phi_{\rm CP} = 0^{\circ}, 30^{\circ}, 60^{\circ}, 90^{\circ}.
\eea
\begin{figure}[hbt]
\includegraphics*[scale=0.4]{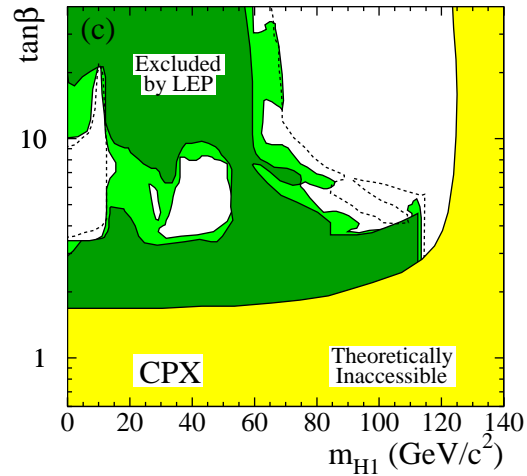}
\caption{\label{lephole} The LEP exclusion region taken from
~\protect\cite{Schael:2006cr}. The light green (medium grey) and 
dark green (dark grey) regions are the exclusion regions
in the $\tan \beta$ -- $m_{H_1}$  plane, at $95$ \%CL and $99.7$ \%CL
respectively.  More conservative of the two theoretical
calculations was used at each point in the parameter space in this figure.
For details such as values of $m_t$ used  and the dependence of the exclusion 
region on them, see 
Ref.~\protect\cite{Schael:2006cr}. The dashed lines indicate the boundaries 
of the regions expected to be excluded at $95 \%$ CL.}
\end{figure}
Combining the results of Higgs searches from ALEPH, DELPHI, L3 and OPAL,
the authors in Ref.\cite{Carena:2002bb,Abbiendi:2004ww,Schael:2006cr,Accomando:2006ga} 
have provided
exclusion regions in the $M_{H_1}$--$\tan \beta $ plane as well as
in the $M_{H^+}$--$\tan \beta $ plane.  While the exact exclusion regions 
differ somewhat in different  analysis
 \cite{Carena:2002bb,Abbiendi:2004ww,Schael:2006cr,Accomando:2006ga} 
and depend on the value of the top quark mass $m_t$, as well as the
programs used to compute the Higgs masses in terms of the model 
parameters, 
they  all show that for phases
$\Phi_{\rm CP} = 90^{\circ} $ and $60^{\circ}$ LEP cannot exclude the
presence of a light Higgs boson at low $\tan \beta$, mainly because of the
suppressed $H_1 ZZ$ coupling. 
\begin{figure*}[thb]
\includegraphics*[scale=0.6]{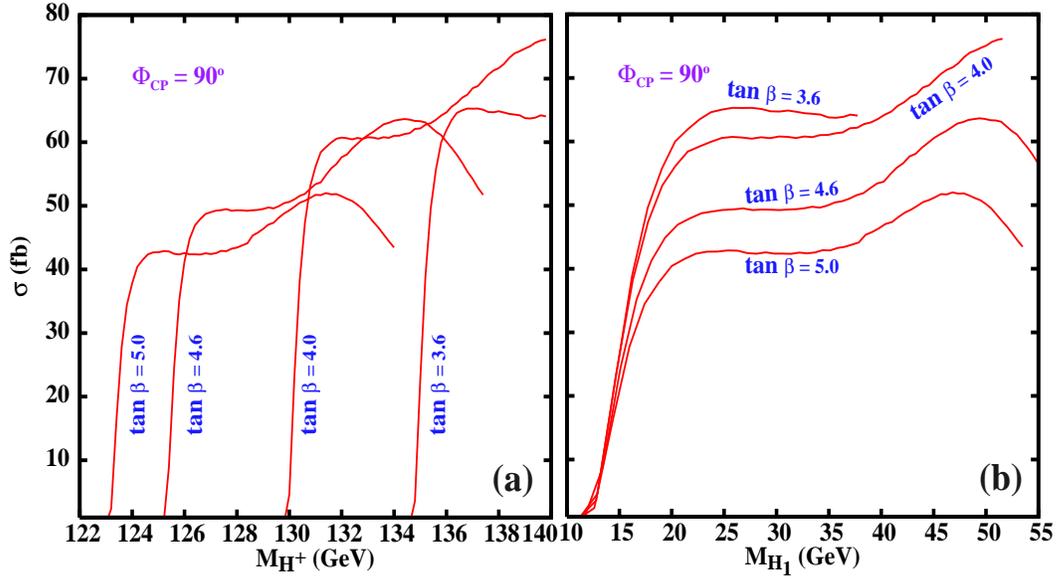}
\caption{\label{csec}
Variation of the  cross-section with $M_{H^+}$
for four values of $\tan\beta =3.6,4,4.6 $ and $5$ is shown in the left panel,
for the CP-violating phase $\Phi_{\rm CP} =   90^{\circ}$.  These  numbers
should be multiplied by $\sim 0.5$ to get the signal cross-section to take
into account the b--tagging efficiency. $M_t, M_W$ mass window cuts have been
applied\protect\cite{Ghosh:2004cc}.}
\end{figure*}
\begin{figure*}[hbt]
\includegraphics*[scale=0.6]{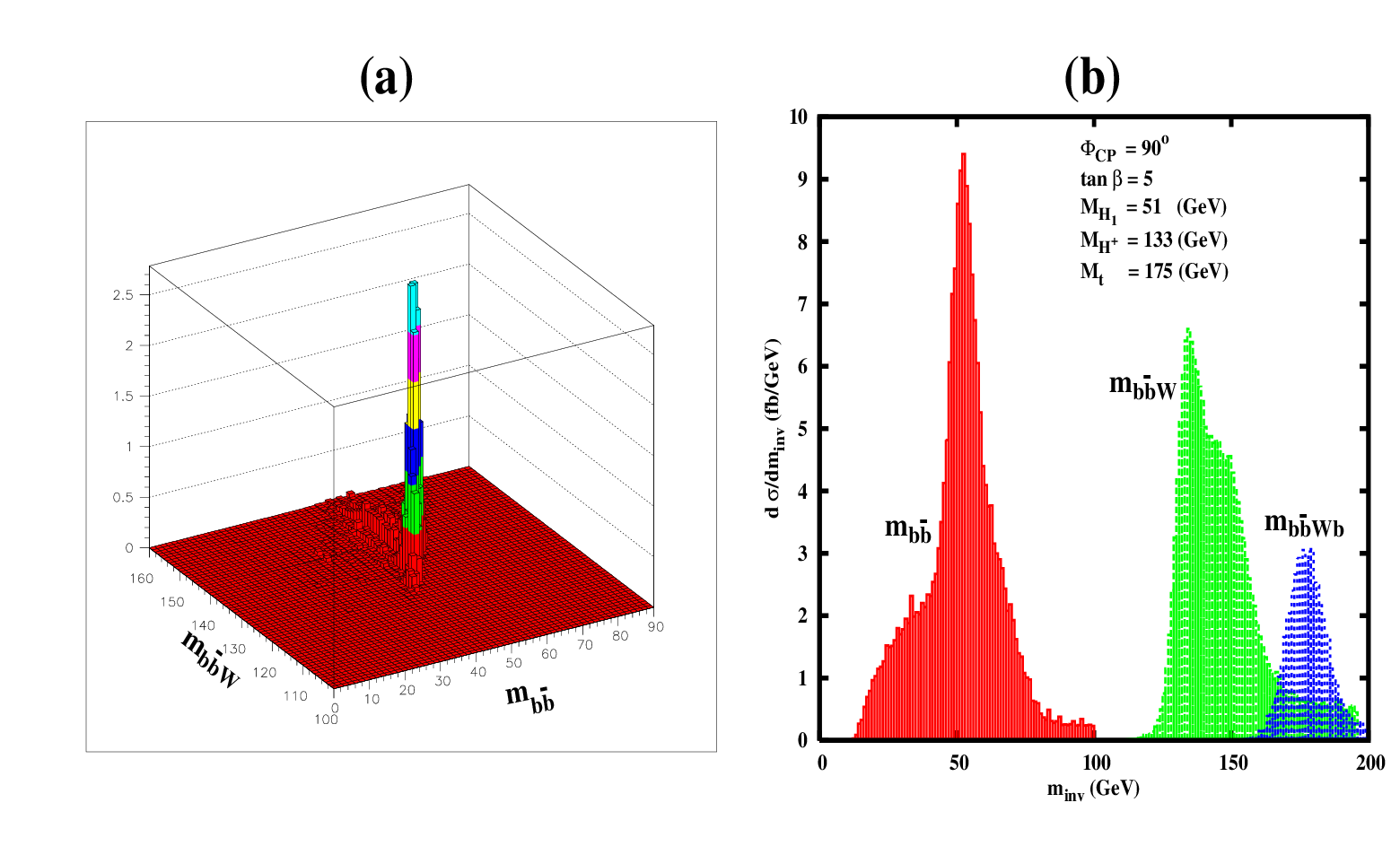}
\caption{\label{corr}
Clustering of the $b\bar b, b\bar bW $ and $b\bar b b W$ invariant
masses. $(a)$ three-dimensional plot for the correlation between
$m_{b\bar b} $ and $m_{b\bar b W}$ invariant mass distribution.
$(b)$ $m_{b\bar b}, m_{b\bar b W}$ and $m_{b\bar b W b}= M_t$
invariant mass distributions for $\Phi_{\rm CP} = 90^{\circ}$.
The other MSSM parameters are $\tan\beta = 5, M_{H^+} = 133$ GeV,
corresponding to a light neutral Higgs $H_1$ with mass $M_{H_1} = 51 $ GeV.
$M_t, M_W$ mass window cuts have been applied\protect\cite{Ghosh:2004cc}.}
\end{figure*}

The analysis of 
Ref.~\cite{Carena:2002bb,schumacher}
further shows that in the same region the $H_1 t \bar t$
coupling is suppressed as well.  Thus  this particular region in the
parameter space can not be probed either at the Tevatron where  the
associated production $W/Z H_1$ mode is the most promising one; neither can
this be probed at the LHC as the reduced $t \bar t H_1$ coupling suppresses
the inclusive production mode and  the associated production modes
$W/Z H_1$ and $t \bar t H_1$, are suppressed as well. This region of
Ref.~\cite{Carena:2002bb} corresponds to $\tan \beta \sim 3.5-5,
M_{H^+}\sim 125-140 $ GeV,~ $ M_{H_1} \stackrel{<}{{}_\sim} 50 $ GeV
and $\tan \beta \sim 2-3, M_{H^+}\sim 105-130~{\rm GeV}, M_{H_1}
\stackrel{<}{{}_\sim} 40 $ GeV, for $\Phi_{CP} = 90^\circ$ and $60^\circ$
respectively. In the same region of the parameter space where $H_1 ZZ$
coupling is suppressed, the $H^+ W^- H_1$ coupling is enhanced because these
two sets of couplings satisfy a sum-rule. Further, in the MSSM a light
pseudo-scalar implies a light charged Higgs, lighter than the top quark.
The comment  about possible constraints on a moderately light charged
Higgs from flavour physics, made in the context of the NMSSM applies in this
case as well. 

\begin{table}
\begin{footnotesize}
\caption{\label{phaseninty}
Range of values  for BR ($H^+ \rightarrow H_1 W^+$) and
BR ($t \rightarrow b H^+ $)
for different values of $\tan \beta$ corresponding to the LEP allowed
window in the CPX scenario, for the common phase $\Phi_{\rm CP} = 90^{\circ}$,
along with the corresponding range for the $H_1$ and
$H^+$ masses. The quantities in the bracket in each column give the values
at the edge of the kinematic region where the decay
$H^+ \rightarrow H_1 W^+$ is allowed.}
\begin{tabular}{|c|c|c|c|}
\hline
&&&\\
$\tan\beta $ &$3.6$&4&5\\[3mm]
\hline
&&&\\
${\rm B} (H^+ \to H_1 W^+)$ & $>90$(87.45)&$> 90 $(57.65)
&$>90$(46.57)  \\
(\%)&&&\\ [3mm]
${\rm B} (t \to b H^+)$ & $\sim$ 0.7 & .7 - 1.1
& 1.0 - 1.3  \\
(\%)&&&\\[3mm]
$M_{H^+}$ & $< 148.5$ (149.9) &$< 139$ (145.8) &
$< 126.2$(134) \\
(GeV)&&&\\[3mm]
$M_{H_1}$ & $< 60.62$ (63.56)  &$< 49.51$ (65.4)  
&$< 29.78$(53.49) \\
(GeV)&&&\\[3mm]
\hline
\end{tabular}
\end{footnotesize}
\end{table}
Table~\ref{phaseninty} shows the behaviour of the $M_{H^+}$, $M_{H_1}$ and
the BR ($H^+ \rightarrow H_1 W^+$), for values of $\tan \beta $
corresponding to the above mentioned window in the $\tan \beta$--$M_{H_1}$
plane, of Ref.~\cite{Carena:2002bb}.
It is to be noted here that indeed the $H^\pm$ is light
(lighter than the top) over the entire range, making its production in $t$
decay possible. Further, the $H^\pm$ decays dominantly into $H_1W$, with a
branching ratio larger than $47\%$ over the entire range where the decay is
kinematically allowed, which covers practically the entire parameter range of
interest; viz. $M_{H_1} <  50 $ GeV for $\Phi_{\rm CP} = 90^\circ $. It can be
also seen from the table that the BR($H^\pm \to H_1 W$) is larger than
$90\%$ over most of the parameter space of
interest. So not only that $H^+$ can be produced abundantly in the $t$ decay
giving rise to a possible production channel of $H_1$ through the decay $H^\pm
\to H_1 W^\pm$, but this decay mode will be the only decay channel to see
a light ($M_{H^\pm} < M_t)$ $H^\pm$. The traditional decay mode of
$H^\pm \to \tau \nu$ is suppressed by over an order of magnitude and thus will
no longer be viable. Thus  the process
$$
p \bar p \rightarrow t(\rightarrow b H^+\rightarrow b W^+ H_1)
\bar t(\rightarrow b W^-\rightarrow b l\nu_l/q \bar q'),$$ with the $H_1$ further
decaying into a $b \bar b$ pair and the $W^+$ decaying into a
$l \nu_l (q \bar q ')$ pair 
will allow a probe of both the light
$H_1$  {\bf and } a light $H^\pm$ in this parameter window in the
CP-violating MSSM in the CPX scenario.

We have investigated the signal over the entire parameter range of the
'hole' in the $\tan \beta$-- $m_{H_1}$ plane of the Figure~\ref{lephole}. We
show our results in Figures~\ref{csec} and \ref{corr},  for common \cp phase 
$\Phi_{CP} = 90^{\circ}$. 
As can be seen from the Figure~\ref{csec} the largest signal
cross-section  case is $\sim 38$ fb and the
signal cross-section is $\gsim  20$  fb for $M_{H_1} \gsim 15$ GeV.
It is clear from the right panel of the Figure~\ref{corr},
that there is simultaneous clustering in the  $m_{b\bar b}$ distribution around
$\simeq M_{H_1}$ and in the $m_{b \bar b W}$ distribution around $M_{H^\pm}$.
It should be mentioned here that the combinatorial background has already been
included in the inclusive $b \bar b$ and $b \bar b W$ invariant mass
distributions plotted in right hand panel in Figure~\ref{corr} whereas the three
dimensional plot showing the correlation does not include this.
The clustering feature can be used to distinguish the signal over the
standard model background.  Technically the most useful in this are the mass 
window cuts ($M_W \pm 15 $ GeV, $M_t \pm 25 $ GeV, $M_{H_1} \pm 15 $ GeV
and $M_{H^\pm} \pm 25 $ GeV on the reconstructed $W, t, H_1$ and $H^\pm$ masses)
employed in the mass reconstruction procedure as described in
Ref.~\cite{Ghosh:2004cc}. As a matter of fact the estimated background 
coming from the QCD production of $t \bar t b \bar b$
once all the cuts (including the mass window cuts) are applied, to
the signal type events is  less than  $0.5$ fb, in spite of a
starting cross-section of 8.5 pb.  The major reduction is brought about by
requiring that the invariant mass of the $b b b W$ be within $25$ GeV of
$M_t$.  Preliminary studies in ATLAS collaboration presented
at Les Houches Workshop~\cite{markus_houch} also show that this background can
be suppressed to negligible levels by similar requirements.
This makes it very clear that the detectability of
the signal is controlled primarily by the signal size. It is also clear from
Figure~\ref{csec} that  indeed the signal size is healthy over the regions of
interest in the parameter space. Thus using this process one can cover the
region of the parameter space in $\cp$ MSSM, in the $\tan\beta -M_{H_1}$ plane
which can not be excluded by LEP-2 and where the Tevatron and the LHC have
no reach via  the usual channels. Note further that this process would be
the only channel of discovery for the charged Higgs-boson $H^\pm$ as well in
this scenario, as the traditional decay mode of $H^\pm  \to \nu \tau $ is
suppressed by over an order of  magnitude.

Note further, that the correlation between a light $\phi$, a moderately 
light $H^\pm$ and the large branching ratio for $H^\pm \rightarrow \phi W^\pm$
follows in the MSSM and in the NMSSM, due to some simple sum rules. 
Hence it may be  a generic feature of any scenario,
which allows a light charged Higgs  in spite of the LEP-II constraints, 
so that $t \rightarrow b H^\pm$ is  not negligible and further 
$H^\pm \rightarrow \phi W^\pm$ is possible, $\phi$ being  a light neutral 
Higgs which dominantly decays into a $b \bar b$ pair. 

\section{Summary}
Thus in conclusion, both in the NMSSM and in the CPV-MSSM the
moderately light charged Higgs that is allowed at moderately
low values of $\tan \beta$, provides interesting and novel phenomenology
at the LHC. It would be interesting to investigate whether the flavour physics
constraints allow a light charged Higgs in the mass range that attains
in these regions of the parameter space in the two cases.

\section{Acknowledgments}
It is  a great pleasure to thank the organizers for this wonderful workshop
and school held in Tehran.  Support of the Indo-French Center for the 
Promotion of Advanced Research under IFCPAR project no. IFC/3004-B/2004 for 
partial support for the research towards some of the results discussed in this 
talk is gratefully acknowledged.

\end{document}